\documentclass{WileyMSP-template}
\usepackage{adjustbox}
\usepackage[flushleft]{threeparttable}
\usepackage{hyperref}
\usepackage{upgreek}
\usepackage{amsmath}
\usepackage{xcolor}
\begin{document}
\newcommand{\mymark}[1]{{\color{red}#1}}
\newcommand{\vmark}[1]{{\color{blue}#1}}

\pagestyle{fancy}
%\rhead{\includegraphics[width=2.5cm]{vch-logo.png}}

\title{Resonant tunneling diode-integrated terahertz transceiver module for wireless communications}

\maketitle

% Author: Please give full first and last names for authors and include * after the name of all corresponding authors

\author{Weijie Gao*}
\author{Nguyen H. Ngo}
\author{Daiki Ichikawa}
\author{Mingxiang Li}
\author{Yuta Inose}
\author{Yuki Morita}\\
\author{Hidemasa Yamane}
\author{Yoshiharu Yamada}
\author{Shuichi Murakami}
\author{Yosuke Nishida}\\
\author{Tadao Nagatsuma}
\author{Withawat Withayachumnankul}
\author{Masayuki Fujita}

% Dedication

%R\dedication{Optional dedication here. If no dedication is required, please leave blank}

% Affiliations: Please provide adacemic titles (Prof. or Dr.) for all authors where applicable, and include an institutional email address for all corresponding authors
\begin{affiliations}
W. Gao, N. H. Ngo, D. Ichikawa, Y. Inose, Y. Morita, T. Nagatsuma, M. Fujita\\
Graduate School of Engineering Science, The University of Osaka, 1-3 Machikaneyama, Toyonaka, Osaka 560-8531, Japan\\
Email Address: gao.weijie.es@osaka-u.ac.jp

M. Li, W. Withayachumnankul\\
Terahertz Engineering Laboratory, Adelaide University, SA 5005, Australia

H. Yamane, Y. Yamada, S. Murakami\\
Osaka Research Institute of Industrial Science and Technology, 2-7-1 Ayumino, Izumi, Osaka 594-1157, Japan

Y. Nishida\\
ROHM Co. Ltd., Kyoto 615-8585. Japan

T. Nagatsuma\\
Graduate School of Science, The University of Tokyo, 7-3-1 Hongo, Bunkyo, Tokyo 113-0033, Japan

\end{affiliations}

% Keywords: Please provide a minimum of three and a maximum of seven keywords, separated by commas

\keywords{Resonant tunnling diode, terahertz transceiver module, terahertz silicon photonics, terahertz communications, 3D-printed lens}

% Abstract should be written in the present tense and impersonal style (i.e., avoid we), and be at most 200 words long
\begin{abstract}
Terahertz bands are essential for next-generation wireless communications, offering ultra-broad bandwidth and unprecedented data throughput. However, realizing compact, low-cost, broadband, and efficient terahertz transceiver modules remains challenging. Conventional modules that rely on metallic hollow-waveguide or silicon-lens packaging suffer from signal loss, bulkiness, and fabrication complexity. Here, we propose a compact terahertz wireless transceiver module enabled by a resonant tunneling diode (RTD) integrated with a photonic–electronic antenna chain. The RTD, grown on an InP substrate, is coupled to a modified Vivaldi antenna functioning as a broadband mode converter to an all-silicon effective-medium-clad dielectric waveguide, which terminates with a rod antenna directly interfaced with a three-dimensional (3D)-printed cyclic olefin copolymer elliptical lens. This configuration enables ultra-broadband and highly directive free-space radiation without additional matching networks or anti-reflection coatings. Encased in a low-cost 3D-printed polylactic acid package, the module attains realized gains of 28–33 dBi for the $E_{11}^x$ mode and 30–33 dBi for the $E_{11}^y$ mode across 220–330 GHz. As a receiver, the module exhibits a noise voltage density down to $5.6\times10^{-9}$~$\rm{V}/\sqrt{\rm{Hz}}$, a minimum noise equivalent power of 1.8~$\rm{pW}/\sqrt{\rm{Hz}}$, and an average responsivity of 6.8~kV/W in the 300-GHz band under amplified detection. It supports error-free transmission (bit error rate (BER) less than $10^{-11}$) up to 30 Gbit/s for on-off keying (OOK) modulation and 80~Gbit/s (BER $<3.8\times10^{-3}$ below hard-decision forward error correction limit) for 16-QAM modulation over 10 cm, and enables real-time uncompressed high-definition video streaming over 1~m. As a transmitter, the module is demonstrated to support OOK transmission at 332 GHz, achieving error-free transmission up to 12~Gbit/s. These results establish a compact, lightweight, and multifunctional terahertz transceiver architecture, highlighting the potential of RTD-based photonic–electronic integration for 6G and beyond wireless frontends.
\end{abstract}

% Text: Please use section headings and subheadings as specified below. For communications, all section headings apart from Experimental Section should be removed
% Please make the first reference to a display item bold: \textbf{Figure 1}
% Do not abbreviate Figure, Equation, etc.; display items are always singular, i.e., Figure 1 and 2.
% Equations are always singular, i.e., Equation 1 and 2, and should be inserted using the {equation} environment, not as graphics
% Please do not use footnotes in the text, additional information can be added to the Reference list.

\section{Introduction}
Terahertz communications have attracted extensive attention as a key enabler for future 6G wireless networks, to provide ultra-broadband links that surpass the fundamental limits of millimeter-wave technologies~\cite{nagatsuma2016,shafie2022,balzer2022}. A critical element of any terahertz communication system is the terahertz frontend~\cite{carlowitz2023}, which serves as the interface between the backend circuitry and free space. Typically, a terahertz frontend consists of a transmitter and/or receiver, waveguide, antenna, and packaging, all designed to maximize efficiency and bandwidth, while enabling system-level integration~\cite{rebeiz2002,dyck2019}. For terahertz-wave generation and detection, extensive efforts have been made in both photonic~\cite{nagatsuma2025} and electronic domains~\cite{suzuki2024}. Yet, differences in fabrication technologies make it difficult to integrate transmitters and detectors on a common platform~\cite{sengupta2018,withawat2018}, often requiring separate hollow-waveguide packages that result in bulky and costly system architectures~\cite{dittmer2023}.

Resonant tunneling diodes (RTDs)~\cite{suzuki2024,chang1974,diebold2016}, consisting of a quantum well sandwiched between two thin barriers exhibit a negative differential conductance (NDC) region in the current-voltage ($I$-$V$) characteristics as a result of quantum tunneling~\cite{tsu1973}. Remarkably, RTDs can operate as both transmitters and receivers at room temperature simply by varying the applied bias voltage. As terahertz oscillators,  RTDs have demonstrated fundamental oscillation frequencies up to 1.98~THz~\cite{asada2021} and output power exceeding 1~mW in the 600--700~GHz range using a single device~\cite{suzuki2024}. Further improvements in both oscillation frequency and output power have been achieved by exploiting harmonic generation~\cite{orihashi2005,yoshida2024} and RTD array configurations~\cite{Koyama2022,han2023}. Due to their strong non-linearity, RTDs can also function as a high-sensitivity direct detectors~\cite{cimbri2022} and self-oscillating mixers through injection-locking~\cite{takida2020,nishida2019}. Using RTDs as both transmitter and coherent receiver, error-free wireless transmission at 30~Gbit/s has been achieved at the 300~GHz band~\cite{nishida2019}. The current record for data rate with RTDs as receivers is 68~Gbit/s for wireless~\cite{webber2023} and 100~Gbit/s for on-chip communications~\cite{ichikawa2025} at the 300~GHz band using multi-level modulations. Most recently, wireless transmission at 860~GHz has been demonstrated using RTDs as both transmitter and receiver, marking a frequency milestone for all-electronic terahertz transceivers~\cite{ngo2025}.

Typically, the RTD oscillator is integrated with a planar antenna such as bow-tie~\cite{diebold2016,oshiro2022,webber2023,webber20218k}, dipole~\cite{tsuruda2020,headland2022rtd}, slot~\cite{oshima2017,van2020,van2022}, split ring resonator~\cite{yu2021h,iwamatsu2021,han2023}, or patch array antenna~\cite{li2023rtd}, to enable efficient power extraction and frequency stabilization. However, these planar antennas are physically small with a limited effective aperture, which restricts their high radiation gain capability. Consequently, they are often positioned at the focal plane of collimating reflector or lens to enhance far-field performance~\cite{rebeiz2002}. In particular, silicon  lenses, which have a dielectric constant close to that of the the InP substrate used for RTDs, act as synthetic infinite dielectrics, effectively suppressing substrate mode losses while improving radiation directivity~\cite{rebeiz2002,li2023rtd,diebold2015}. Commonly employed lens geometries include elliptical~\cite{wu2001,gulan2013,montero2018,dyck2019,derat2020,gashi2022,ziegler2024}, hyperhemispherical~\cite{park2013,dancila2016}, hemispherical, and extended hemispherical profiles~\cite{filipovic2002,godi2005,llombart2011,nguyen2012,konstantinidis2017}. Despite their excellent optical properties, silicon (Si) lenses are costly and typically require anti-reflection coatings at the air-dielectric interface to minimize reflection losses~\cite{gashi2022,nguyen2009}, further increasing fabrication complexity and cost. To reduce the cost, high-density-polyethylene (HDPE) lenses with a much lower dielectric constant have been introduced~\cite{diebold2015}. However, an impedance-matching layer between the HDPE lens and the diode substrate is required, which limits the achievable bandwidth. Moreover, the large physical size mismatch between the RTD chip and the lens makes the radiation patterns highly sensitive to the alignment~\cite{diebold2015}, requiring precise assembly~\cite{tsuruda2020}.

To improve integrability and minimize transmission-line-induced loss, all-dielectric waveguides have been introduced to interface with RTDs through heterogeneous and hybrid integrations~\cite{headland2022rtd,yu2019,yu2021hybrid,kawamoto2023rtd,ngo2023rtd}. However, the inherent mode mismatch between the diode and waveguide necessitated the use of a mode converter~\cite{rebeiz2002}. The initial implementation employed a high-resistivity silicon photonic crystal (PC) waveguide~\cite{tsuruda2015} coupled with an RTD-integrated Vivaldi antenna, achieving a $90\%$ coupling efficiency~\cite{yu2019,yu2021hybrid}. A silicon taper at the waveguide end provided input/output (I/O) interfacing to hollow waveguide and free space, enabling 32~Gbit/s error-free transmission at 350~GHz via wired interconnection~\cite{yu2019}. However, the strong in-band dispersion of the PC structure limited the bandwidth to below $10\%$. To extend bandwidth, a slot dipole antenna was introduced to backside couple to an effective-medium(EM)-clad dielectric waveguide~\cite{gao2019,gao2020,headland2020unclad}, leading to a fractional bandwidth of $28\%$~\cite{headland2022rtd}. Although this reduced the chip size to match the waveguide core, the reliance on gravity for chip support required high-precision alignment and lacked structural robustness. To overcome this limitation, an EM-clad dielectric waveguide combined with a corrugated Fermi-Dirac antenna was developed, significantly improving bandwidth and coupling efficiency~\cite{kawamoto2023rtd}. A silicon cap on the RTD chip  mitigated reflection losses caused by thickness-induced mode and impedance mismatch, yielding a tenfold signal-to-noise ratio (SNR) improvement compared to the PC waveguide-based configuration~\cite{kawamoto2023rtd}. To further simplify integration, a modified Vivaldi antenna was later introduced, enhancing directivity, suppressing out-of-plane radiation, and eliminating the need for silicon cap~\cite{ngo2023rtd}. For system-level interfacing, the RTD-waveguide assembly was mounted in a metallic hollow-waveguide package, with I/O coupling via the silicon taper and a hollow waveguide~\cite{ngo2023rtd,kawamoto2021}. A horn antenna could be directly attached for free-space radiation. While compatible with conventional terahertz communications setups, this approach suffers from metallic loss, high fabrication cost, and limited scalability.  

In this work, we propose an RTD-based transceiver module that integrates a modified Vivaldi antenna as a mode converter with an EM-clad rod antenna, which is directly coupled to a 3D-printed cyclic olefin copolymer (COC) elliptical lens for highly directive free-space radiation. The module is encased by a 3D-printed polyactic acid (PLA) package, forming a compact and low-loss frontend. Such a design strategy eliminates the need for complex matching networks typically required for lens coupling in transistor-based circuits~\cite{dyck2019}. Unlike conventional RTD modules, this approach requires no matching layer~\cite{diebold2015} or anti-reflection coating~\cite{li2023rtd} for lenses, while offering ultra-broadband operation. The 3D printed dielectric package further enhances efficiency by removing the package-induced ohmic losses, while significantly reducing fabrication complexity and cost~\cite{song2017}. The proposed architecture can be expanded as a multifunctional terahertz radio frequency (RF) frontend, supporting the integration of passive components such as filters~\cite{gao2021,chen2025}, frequency-~\cite{headland2021gratingless} and polarization-multiplexers~\cite{gao2025ultra}, and beamformers~\cite{panisa20231,gao2025beam} on the same EM-clad waveguide platform. Section II presents the proposed design, while Section III demonstrates communications experiments with the module as a transceiver, followed by the conclusion given in Section IV. 
\begin{figure}[!tb]
	\centering
	\includegraphics[width=0.85\linewidth]{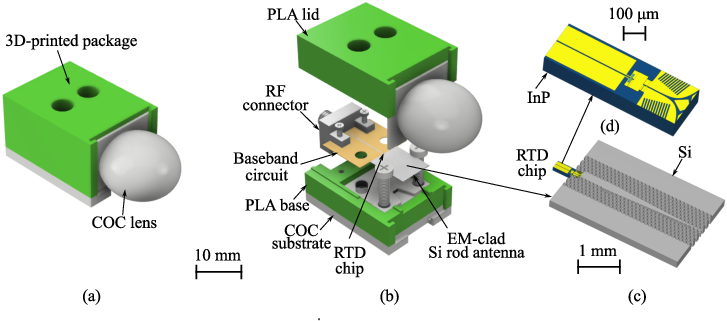}
	\caption{Schematic of the proposed RTD-based terahertz transceiver module. (a) Assembled module. (b) Disassembled module. Magnified view of the (c) EM-clad tapered rod antenna and (d) RTD chip. A 3D-printed COC substrate is inserted under the base of the package to support the RTD and dielectric antenna, as shown in (b).}
	\label{fig:module}
\end{figure}
\section{Design concepts}
\subsection{Overview}
The schematic of the proposed transceiver module is shown in Fig.~\ref{fig:module}. The module is packaged in a 3D-printed PLA enclosure attached with a COC elliptical lens for free-space coupling as shown in Fig.~\ref{fig:module}(a). Inside, the RTD chip, fabricated on an InP substrate, is integrated with a modified Vivaldi antenna for mode conversion to a Si EM-clad dielectric tapered rod antenna that is directly coupled to the COC lens, as illustrated in ~\ref{fig:module}(b). The RTD chip is wire-bonded to a coplanar-waveguide-based baseband circuit, which is connected to a 2.92-mm coaxial connector for DC bias and intermediate-frequency (IF) signal I/O. The RTD chip and the dielectric antenna shown in Fig.~\ref{fig:module}(c) are mounted on a COC supporting substrate to minimize the dielectric loss. By simply adjusting the bias voltage, the module can operate as either a transmitter or receiver, enabling flexible system functionality. The entire structure is established on an electronic-photonic antenna chain to efficiently deliver or collect terahertz waves. The modified Vivaldi antenna expands the deep-subwavelength RTD mode to match the waveguide mode, while the rod antenna provides a smooth transition from the dielectric waveguide to the lens with the guided mode gradually leaked. The lens collimates the leaked wave for significant gain enhancement. Owing to the broadband nature of each antenna stage, the module supports ultra-broadband operation, concatenating the WR-3.4 (220--330~GHz) and WR-2.8 (260--400~GHz) bands with a fractional bandwidth of 54.5$\%$. Furthermore, the EM-clad waveguide platform allows the integration of various on-chip routing and functional components, whose interoperability enables a compact and multifunctional terahertz RF frontend.

\subsection{Modified Vivaldi antenna}
\begin{figure}[!tb]
	\centering
	\includegraphics[width=0.7\linewidth]{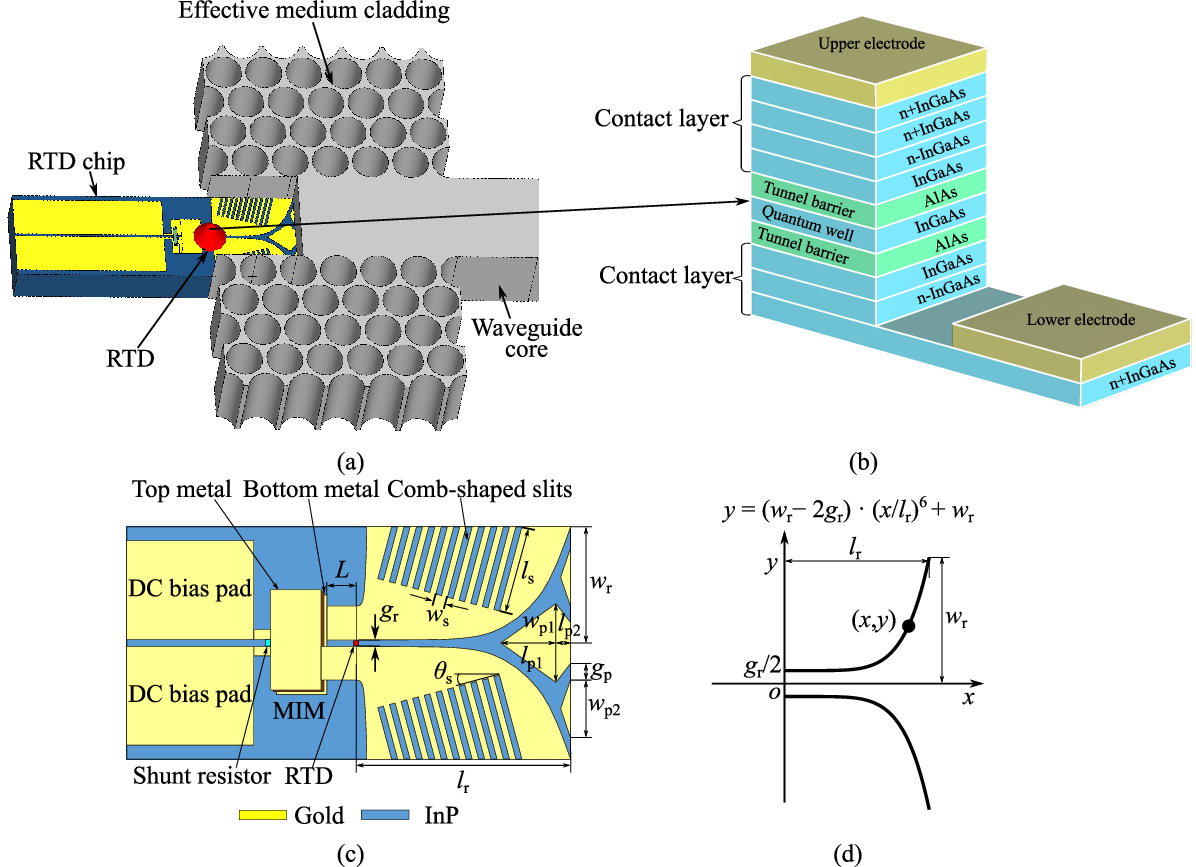}
	\caption{Schematic of RTD chip integrated with EM-clad dielectric waveguide. (a) Perspective view of the integration with RTD chip and the dielectric waveguide. (b) Eptitaxial quantum-well structure of RTD mesa~\cite{diebold2016}. (c) Design of RTD chip. (d) Profile of the modified Vivaldi antenna. The key dimensions of the RTD chip are given as: $L=20~\upmu$m, $g_{\rm{r}}=6~\upmu$m, $w_{\rm{s}}=12~\upmu$m, $l_{\rm{s}}=87.2~\upmu$m, $\theta_{s}=15^{\circ}$, $w_{\rm{p1}}=80~\upmu$m, $l_{\rm{p1}}=55~\upmu$m, $l_{\rm{p2}}=15~\upmu$m, $w_{\rm{p2}}=60~\upmu$m, $g_{\rm{p}}=16~\upmu$m, $w_{\rm{r}}=116~\upmu$m, $l_{\rm{r}}=212~\upmu$m. The duty cycle of the slit array is 50\%. In CST simulations, the dielectric waveguide shown in (a) is terminated with a tapered structure to couple to the feeding hollow waveguide~\cite{gao2019}.}
	\label{fig:RTD}
\end{figure}

Figure~\ref{fig:RTD}(a) illustrates the integration of RTD chip and EM-clad dielectric waveguide. The RTD chip is fabricated on a 100-$\upmu$m-thick semi-insulating  InP substrate with a relative permittivity of 12.6. As shown in Fig.~\ref{fig:RTD}(b), its epitaxial structure comprises a InGaAs quantum well sandwiched between two thin AlAs tunneling barriers, forming a double-barrier quantum well that exhibits negative differential conductance (NDC) in the $I$-$V$ characteristics and enables spontaneous terahertz radiation. As illustrated in Fig.~\ref{fig:RTD}(c), the RTD is positioned between a modified Vivaldi antenna, serving as a mode converter, and a metal-insulator-metal (MIM) reflector that suppresses backward terahertz wave leakage~\cite{yu2019,yu2021hybrid}. In addition, the MIM mirror is also used to separate direct current (DC) and RF components and functions as a low-pass filter for communications. To further suppress the undesired lower-frequency parasitic oscillations, a shunt resistor with $R_{\rm{s}}=30~\Omega$ is integrated after the MIM capacitor. The  transmission lines between the MIM and RTD severing as a shorted tuning stub is carefully designed, where its inductance together with the RTD capacitance determines the oscillation frequency of the chip. Two electrode pads are provided for wire bonding to the baseband circuit for signal I/O.

A tapered slot antenna based on a modified Vivaldi structure is employed due to its ultra-broad bandwidth. In a conventional Vivaldi antenna, as the slot exponentially flares, the local impedance matches that of free space, where high-frequency components radiate earlier along the taper and low-frequency components radiate near the aperture~\cite{gibson1979}. The flare width governs the lower frequency limit and the coupling efficiency with dielectric waveguide, typically designed as half the guided wavelength of the lowest operating frequency~\cite{bai2011}. A wider flare extends bandwidth but increases slot length, leading to higher metallic loss and a larger footprint, whereas a narrower flare introduces modal mismatch and reduced coupling efficiency~\cite{yu2019}. To balance bandwidth, loss, coupling efficiency, and structural compactness, the exponential profile of the conventional Vivaldi is replaced by a polynomial function, as shown in Fig.~\ref{fig:RTD}(d). In this design, the total flare width is constrained to match that of the dielectric waveguide operating in a fundamental mode, while the tapered slot length is shortened to minimize metallic loss and ensure a smooth modal evolution. In this case, a polynomial function $f(x)=x^n$ is adopted as a basic function, where $n$ is optimized as 6 to satisfy the above criteria, and the dimensions are normalized to the antenna length as shown in Fig.~\ref{fig:RTD}(d). Compared to the exponential profile, the polynomial function offers a greater flexibility in controlling the flare rate while maintaining a compact antenna geometry. 

\begin{figure}[!htb]
	\centering
	\includegraphics[width=0.7\linewidth]{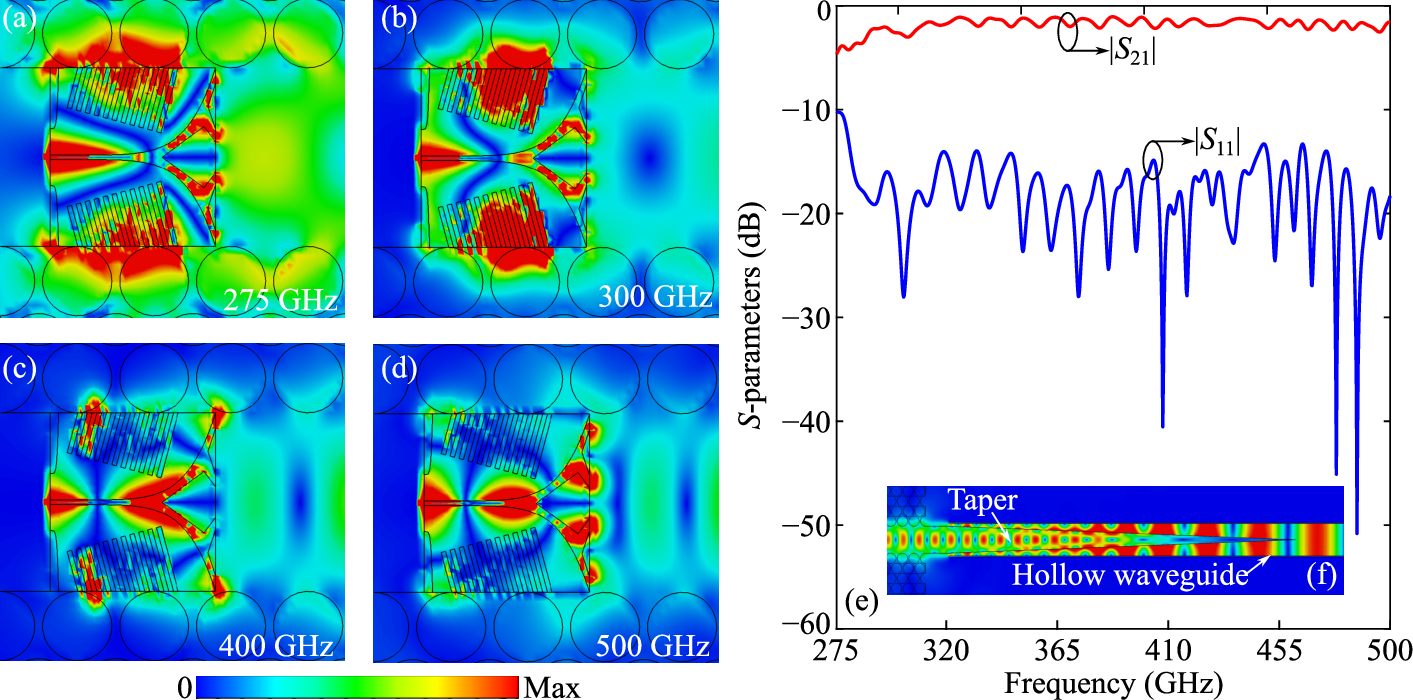}
	\caption{Simulated results of RTD chip integrated with the dielectric waveguide. $E$-field distributions at (a) 275~GHz, (b) 300~GHz, (c) 400~GHz, and (d) 500~GHz. (e) Scattering ($S$)-parameters. (f) Coupling between the tapered structure and the hollow waveguide. All the $E$-field distributions are in linear scale and normalized by the same factor.}
	\label{fig:S21RTD}
\end{figure}

To further enhance the bandwidth, periodic comb-shaped slits are introduced along the antenna arms~\cite{teni2013,moosazadeh2015,moosazadeh2016}. These slits increase current density, which effectively lengthens the electrical path, thereby lowering operation frequencies and broadening the bandwidth~\cite{teni2013,moosazadeh2015,fei2011}. This behavior is evident in the simulated electric ($E$)-field distributions shown in Figs.~\ref{fig:S21RTD}(a)-(d), where the lower-frequency components occupy a larger portion of the slit region and exhibit a stronger coupling with the slits. At higher frequencies, the spherical wavefront in the flare becomes more pronounced as the effective aperture increases. However, this can cause phase reversal along the antenna edges, resulting in gain degradation and elevated sidelobes~\cite{nassar2015}. To mitigate this effect, a diamond-shaped parasitic element is introduced at the center of the aperture~\cite{li2016}, dividing the tapered slot into two smaller apertures with coupled fields, as shown in Figs.~\ref{fig:S21RTD}(a)-(d). This design suppresses off-axis radiation caused by phase errors due to direct coupling between the antenna arms~\cite{nassar2015}, while maintaining a compact footprint. Triangular loads at the ends of each sub-aperture further improve directivity and impedance matching. As a result, the modified Vivaldi antenna achieves enhanced directivity and efficient coupling to the dielectric waveguide across a broad bandwidth with a compact footprint, efficiently concentrating the radiated power into the dielectric waveguide core.

The RTD chip is directly coupled to an EM-clad dielectric waveguide with a thickness of 200~$\upmu$m. Notably, the proposed design can effectively suppress the out-of-plane radiation, eliminating the need for a substrate cap to mitigate the thickness-induced coupling loss, thereby improving integration practicality. To evaluate the transmission coefficient, a silicon tapered structure is attached to the dielectric waveguide and coupled to a hollow waveguide for signal extraction in Fig.~\ref{fig:RTD}(a). The full-wave simulations are conducted by CST Microwave Studio Suit 2025. As shown in Fig.~\ref{fig:S21RTD}(e), simulations confirm broadband performance from 275--500~GHz with an average coupling efficiency above 80$\%$. The corresponding $E$-field distributions at various frequencies, shown in Figs.~\ref{fig:S21RTD}(a)-(d), reveal a smooth energy transition between the tapered slot antenna and the dielectric waveguide. The terahertz waves propagate through the dielectric waveguide and are efficienctly coupled into the receiving hollow waveguide via the silicon taper structure as shown in Fig.~\ref{fig:S21RTD}(f).

\subsection{Lens-coupled rod antennas} 
\subsubsection{Air-clad rod with symmetric elliptical lens} 
\begin{figure}[!tb]
	\centering
	\includegraphics[width=0.8\linewidth]{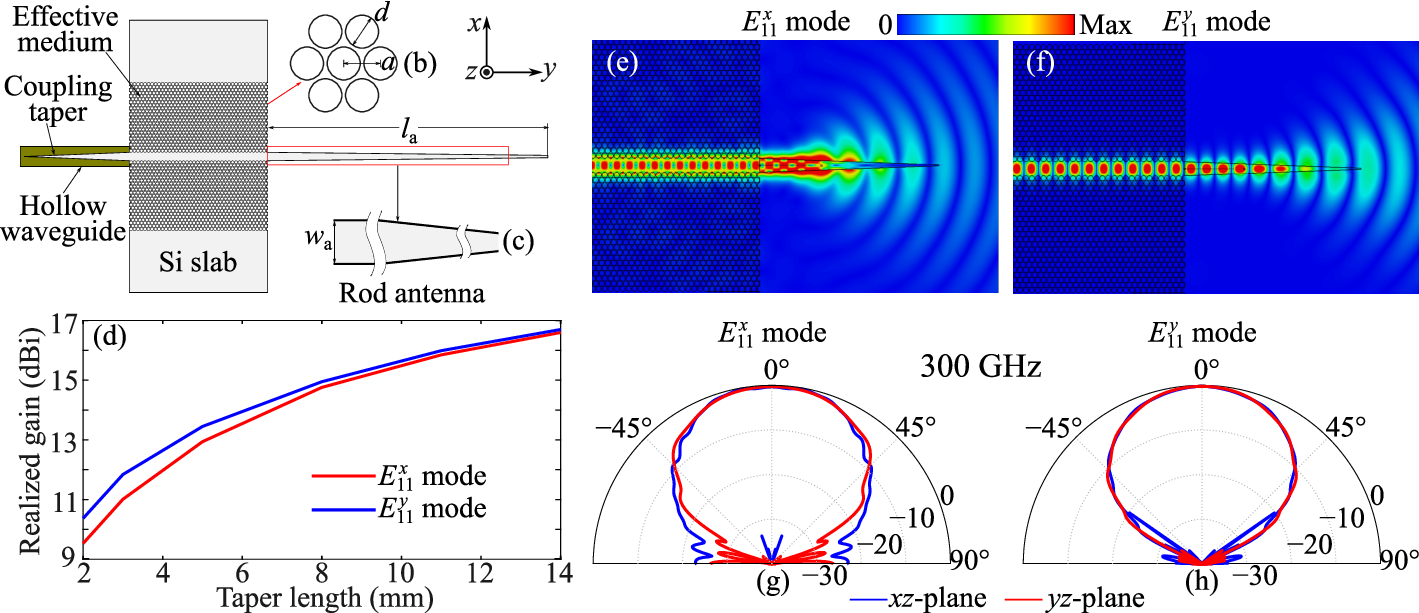}
	\caption{Characteristics of air-clad tapered rod antenna fed by EM-clad waveguide. (a) Schematic of the antenna. Magnified view of (b) the hexagonal lattice of the air-silicon effective medium cladding, and (c) the rod antenna. (d) Simulated realized gain of the rod antenna versus taper length at 300~GHz. Simulated $E$-field distributions for (e) $E_{11}^{x}$ and (f) $E_{11}^{x}$ modes at 300~GHz. Simulated radiation patterns of a 3-mm tapered rod antenna for (g) $E_{11}^{x}$ and (h) $E_{11}^{x}$ modes at 300~GHz.  The loss tangent of silicon adopted in the CST is $3\times10^{-5}$. Here, $a=100~\upmu$m, $d=90~\upmu$m.}
	\label{fig:air_rod_charc}
\end{figure}

For free-space radiation, we first consider a basic model with a single air-clad tapered rod antenna fed by an EM waveguide as shown in Figs.~\ref{fig:air_rod_charc}(a)-(c). A single tapered rod inherently provides broadband and high-gain radiation by gradually leaking the guided mode into free space, as illustrated in Figs.~\ref{fig:air_rod_charc}(d)-(e), forming a large effective aperture as shown in Figs.~\ref{fig:air_rod_charc}(g)-(h), where the leakage frequency increases along the taper~\cite{halliday1947,rivera2019,generalov2014}. This antenna features structural simplicity, low loss, broad bandwidth, and frequency-independent radiation patterns, while supporting two orthogonal polarizations. The gain of the rod antenna is proportional to its radiation length governed by the mode leakage rate, and increasing the taper length can raise the gain as shown in Fig~\ref{fig:air_rod_charc}(f). Forming a rod antenna array~\cite{pousi2009,rivera2017,reese2018,withawat2018rod} can increase the effective aperture and gain further, but requires additional feeding networks or a collimating lens, resulting in greater structural complexity and a fan-beam radiation pattern for planar implementations.
\begin{figure}[!htb]
	\centering
	\includegraphics[width=0.65\linewidth]{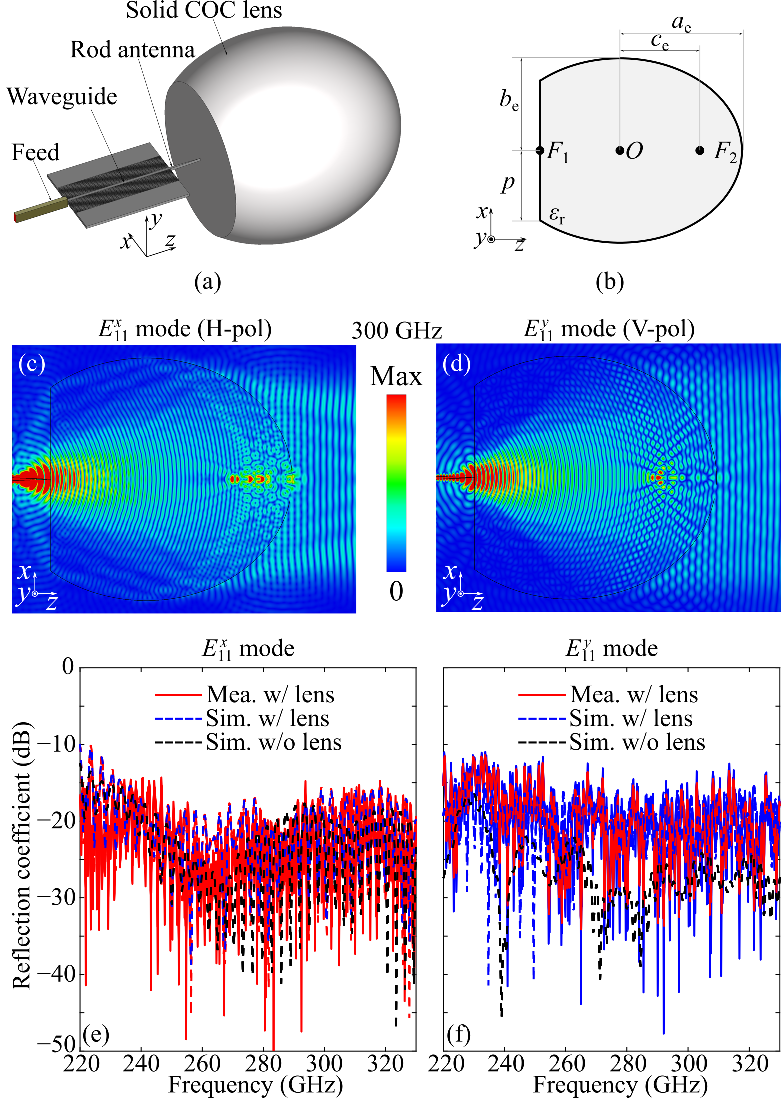}
	\caption{Air-clad tapered rod antenna coupled to an azimuthally symmetric elliptical lens. (a) Perspective view of the antenna architecture. (b) Profile of the elliptical lens. Simulated $E$-field distributions of the lens antenna for (c) $E_{11}^x$ mode with horizontal polarization ($xz$-plane) and (d) $E_{11}^y$ mode with vertical polarization ($yz$-plane) at 300~GHz. Simulated and measured reflection coefficients for (e) $E_{11}^x$ and (f) $E_{11}^y$ modes over WR-3.4 band (220--330~GHz). The reflection coefficients were measured using a Keysight VNA with VDI extenders spanning 220--330~GHz.}
	\label{fig:waveguide}
\end{figure}
\begin{figure}[!tb]
	\centering
	\includegraphics[width=0.9\linewidth]{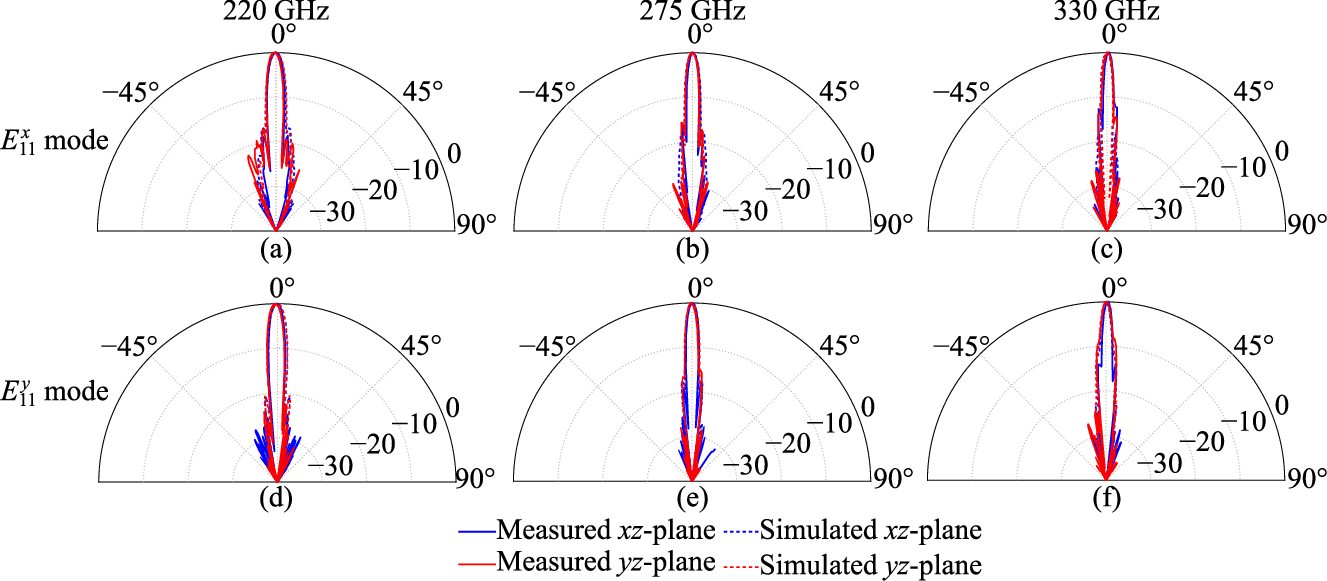}
	\caption{Simulated and measured radiation patterns of the lens-coupled rod antenna at 220~GHz, 275~GHz, and 300~GHz for (a-c) $E_{11}^x$ mode with in-plane polarization and (d-f) $E_{11}^y$ mode with out-of-plane polarization. The measurement setup is described in the Experimental Section.}
	\label{fig:RP1}
\end{figure}
To this end, we introduce a 3D-printed COC elliptical lens~\cite{gao20233d} coupled to the rod antenna to further enhance the gain over a broad bandwidth as shown in Fig.~\ref{fig:waveguide}(a). The length of the rod antenna is set as 3~mm to ensure sufficient mechanical strength while maintaining a reasonably high gain around 10~dBi. The COC material is used mainly because of its exceptionally low loss at terahertz frequencies with a realistic loss tangent approximately $7\times10^{-4}$ and its low relative permittivity of $\epsilon_{\rm{c}}=2.33$~\cite{chung2024}, which eliminates the need for matching layers at both the lens-air and lens-tapered rod interfaces. In this design, we adopt a true elliptical profile to achieve superior performance compared  hemispherical and hyper-hemispherical lenses~\cite{rebeiz2002}. Moreover, 3D printing technology enables precise fabrication of elliptical surface, overcoming the manufacturing challenges associated with conventional machining methods.

The profile of the elliptical lens is shown in Fig.~\ref{fig:waveguide}(b), where its optical dimensions are governed by the following relations~\cite{dyck2019}:  
\begin{align}
	a_{\rm{e}} &=\frac{\epsilon_{\rm{c}}}{\epsilon_{\rm{c}}-1}\cdot p \label{eq1}, \\
	b_{\rm{e}} &=\sqrt{1-\frac{1}{\epsilon_{\rm{c}} }}\cdot a_{\rm{e}} \label{eq2}, \\
	c_{\rm{e}} &=\sqrt{a_{\rm{e}}^2-b_{{e}}^2} \label{eq3}.
\end{align}
Here,  $a_{\rm{e}}$ and $b_{\rm{e}}$ denote semimajor and semiminor axes, respectively, and $c_{\rm{e}}$ is the linear eccentricity of the ellipse. The semilatus rectum $p$ is a key design parameter that determines both the lens size and its focusing performance. It is optimized to achieve an average realized gain exceeding 30~dBi for both orthogonal polarizations over  220--400~GHz, while maintaining a reasonably compact footprint. The optimal value of $p$ is 6.5~mm, allowing the lens to fully capture the lower-frequency radiation with wider beamwidths from the tapered rod antenna. The rod antenna tip, approximated as a point source, is placed at the first focus $F_{1}$. The emitted rays are refracted by the elliptical surface and converge in phase at the second focus $F_{2}$, forming a highly collimated wavefront that propagates as a nearly plane wave, as illustrated in Figs.~\ref{fig:waveguide}(c)-(d). This behavior can be simply understood from the Huygens's principle and the constant sum of optical path lengths between any point $P$ on an ellipse and its two foci ($|PF_{1}|+|PF_{2}|\equiv2a_{\rm{e}}$).

\begin{figure}[!tb]
	\centering
	\includegraphics[width=0.65\linewidth]{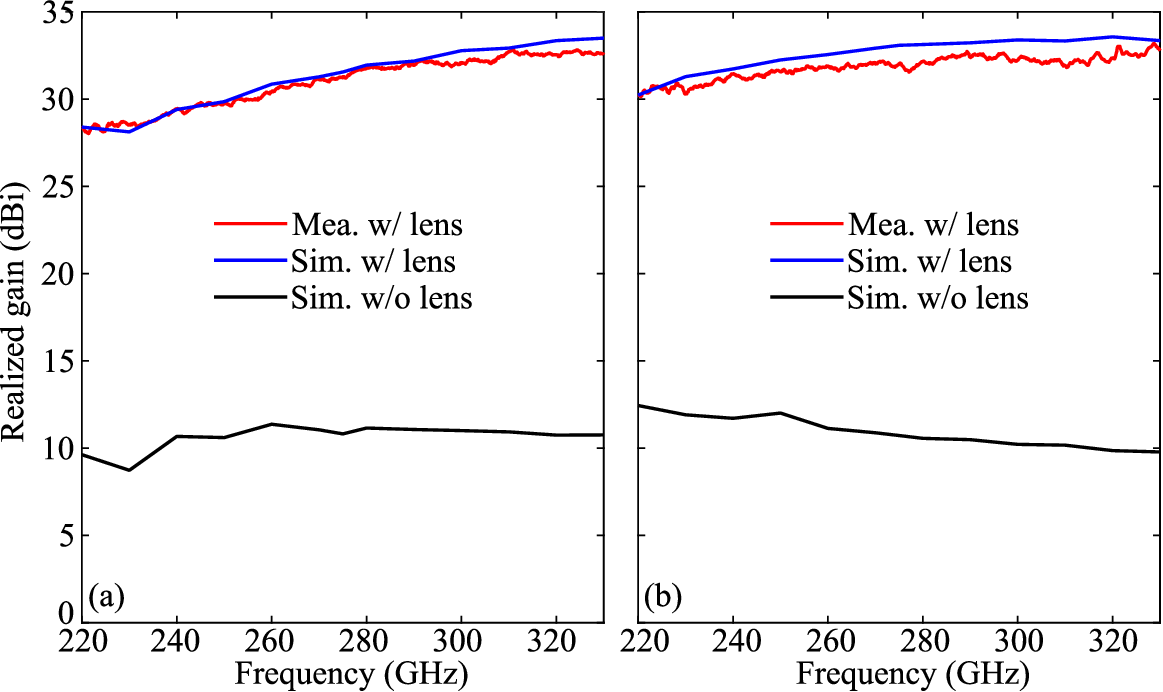}
	\caption{Simulated and measured realized gains of the antennas in Fig.~\ref{fig:waveguide}(a) for (a) $E_{11}^x$ and (b) $E_{11}^y$ modes.}
	\label{fig:gainrod}
\end{figure}
The simulated and measured reflection coefficients for the $E_{11}^x$ and $E_{11}^y$ modes are shown in Figs.~\ref{fig:waveguide}(e)-(f), where the measured results agree well the simulations. For both polarizations, $|S_{11}|$ remains below $-10$~dB across the band. Rapid spectral fluctuations in the measured data are attributed to discontinuities in conjoined waveguide sections and external reflections as demonstrated in the Experimental Section. Compared to the bare rod, the lens-coupled rod exhibits slightly higher reflections due to weak impedance mismatch at the rod-lens interface, though this has a negligible impact on the overall performance. The simulated and measured radiation patterns in Fig.~\ref{fig:RP1} show excellent agreement for both modes. The measured 3-dB angular beamwidths in the $xz$ and $yz$ planes are less than $5^\circ$, producing pencil-like beams highly desired for terahertz communications. As shown in Fig.~\ref{fig:gainrod}, the measured realized gain ranges from 28 to 33~dBi for the $E_{11}^x$ mode and from 30 to 33~dBi for the $E_{11}^y$ mode. The slight reduction from simulations arises mainly from the misalignment between the feeding hollow waveguide and the coupling taper of the dielectric waveguide, which introduces additional coupling loss. Additionally, it is found that small off-axis movement of the rod antenna relative the lens can degrade the gain. Since the radiation beam is highly directive, small alignment errors between the transmitter and receiver can also affect the measured gain. Nevertheless, the lens enhances the rod antenna gain by around 20~dB, while the gain increases with frequency as the effective aperture expands. For the $E_{11}^y$ mode, however, the realized gain of the bare rod antenna decrease at higher frequencies because tighter wave confinement weakens radiation along the taper, reducing the effective aperture. This behavior differs from the $E_{11}^x$ mode due to the mode-dependent dispersion of the effective-medium waveguide~\cite{gao2020}. Notably, the elliptical lens compensates this reduction for the $E_{11}^y$ mode, flattening the gain response, whereas for the $E_{11}^x$ mode, the gain continues to rise with frequency, slightly narrowing the 3-dB gain bandwidth.

\subsubsection{Effective medium-clad rod with asymmetric lens}
\begin{figure}[!b]
	\centering
	\includegraphics[width=0.65\linewidth]{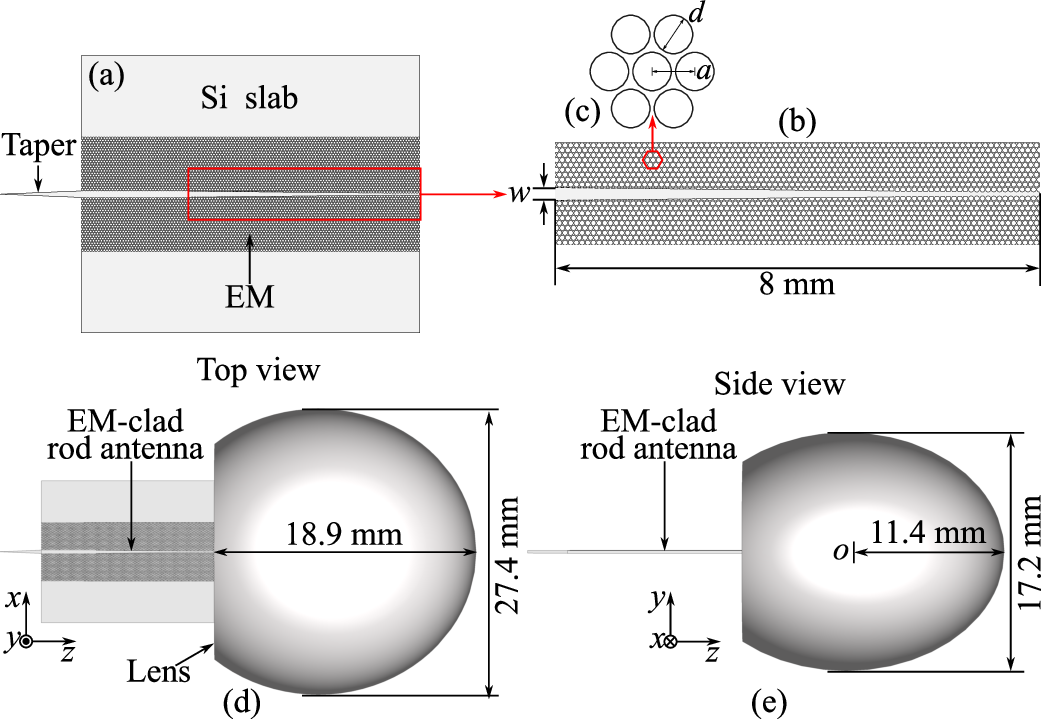}
	\caption{EM-clad tapered rod antenna with an asymmetric elliptical lens. (a) Schematic of the EM-clad tapered rod antenna. Magnified view of the (b) tapered rod surrounded by effective medium in a (c) hexagonal lattice. (d) Top view and (e) side view of the combined antenna architecture. The key dimensions of the air-clad rod are $a=100~\upmu$m, $d=90~\upmu$m, $w=230~\upmu$m. The etching depth is equal to the thickness of the Si slab $h=200~\upmu$m.}
	\label{fig:asymlens}
\end{figure}
\begin{figure}[!tb]
	\centering
	\includegraphics[width=0.65\linewidth]{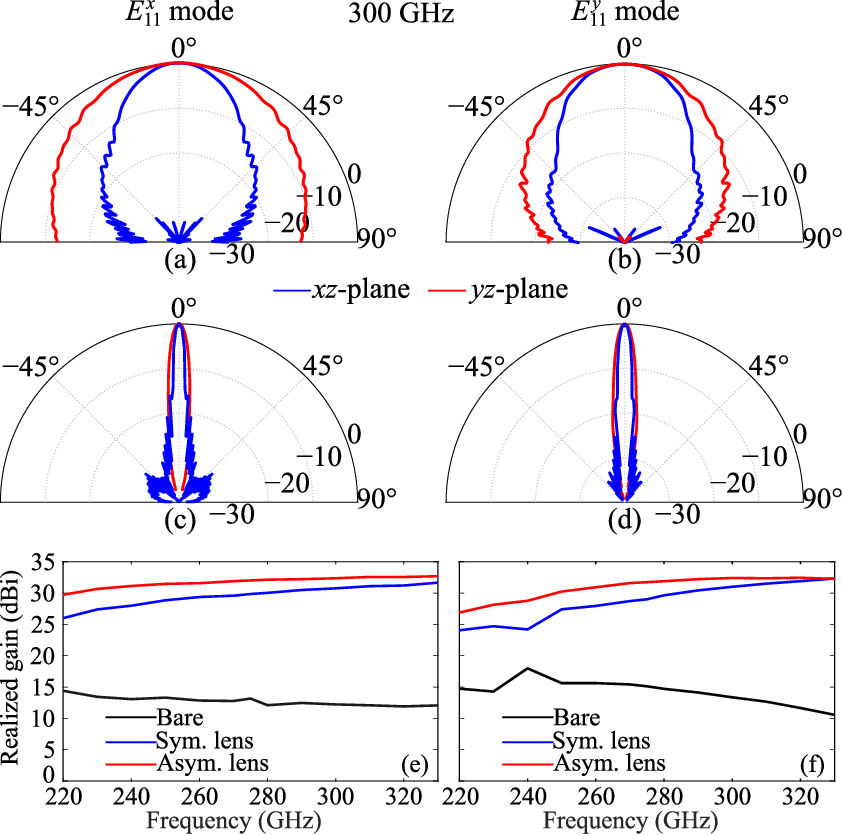}
	\caption{Simulated results of the antennas shown in Fig.~\ref{fig:asymlens}. Radiation patterns of the bare EM-clad rod antenna for (a) $E_{11}^x$ and (b) $E_{11}^y$ modes at 300~GHz. Radiations patterns for the tapered antenna with an asymmetric elliptical lens for (c) $E_{11}^x$ and (d) $E_{11}^y$ modes at 300~GHz. Comparison of realized gains for (e) $E_{11}^x$ and (f) $E_{11}^y$ modes over WR-3.4 band.}
	\label{fig:rpEMrod}
\end{figure}
The air-clad rod-lens antenna discussed above demonstrate excellent performance with a simple design, validating the effectiveness of the proposed strategy. However, several practical challenges remain. First, the tapered rod is mechanically fragile, making it prone to breakage when extended to longer lengths. Second, the point-surface contact between the rod and lens makes the antenna performance highly sensitive to alignment. In addition, gain response between two polarizations diverge more significantly at higher frequencies. To address these issues, an EM-clad tapered rod antenna is introduced as shown in Fig.~\ref{fig:asymlens} (a). Here, the tapered rod is surrounded by in-plane effective medium claddings as illustrated in Figs.~\ref{fig:asymlens}(b)-(c), which enhance mechanical robustness and allow for surface-surface contact with the lens, relaxing alignment tolerances while preserving radiation characteristics of the rod antenna. The simulated radiation patterns for the EM-clad rod antenna are presented in Figs.~\ref{fig:rpEMrod}(a)-(b). The angular beamwidths in the $xz$- and $yz$-plane are unequal, with the narrower $xz$-plane beamwidth resulting from radiation through the EM claddings. Specifically, the evanescent filed of the guided waves gradually leak into the claddings~\cite{gao2020,gao2021} and radiate outward, effectively expanding the in-plane radiation aperture.

However, the fan-beam radiation can reduce the realized gain, while the in-plane EM introduces out-of-plane scattering that becomes more pronounced at higher frequencies. To mitigate these effects, we introduce a modified elliptical lens with asymmetric profiles in the $xz$- and $yz$-plane, as illustrated in Figs.~\ref{fig:asymlens}(d)-(e). Specifically, the semiminor axis of the ellipse in the $xz$-plane is made larger than that in the $yz$-plane to match the wider effective aperture, resulting in nearly identical beamwidths in both planes, as seen in Figs.~\ref{fig:rpEMrod}(c)-(d). Compared with a symmetric lens, this asymmetric design provides a flatter gain response across a wide bandwidth and exhibits notable performance improvement at lower frequencies, as seen in Figs.~\ref{fig:rpEMrod}(e)-(f). Meanwhile, the gain divergence between two orthogonal modes reduces with frequency increasing. Although the $E_{11}^x$ mode is primarily used for RTD integration, maintaining balanced performance between both orthogonal modes is advantageous for future polarization-multiplexed frontends. While the asymmetric lens slightly increases the footprint in the $xz$-plane, the overall structure significantly enhances the radiation performance and mechanical robustness, making it a more suitable configuration for transceiver module integration.

\section{Transceiver module and communications demonstration}
\subsection{Realized transceiver module}
\begin{figure}[!tb]
	\centering
	\includegraphics[width=0.65\linewidth]{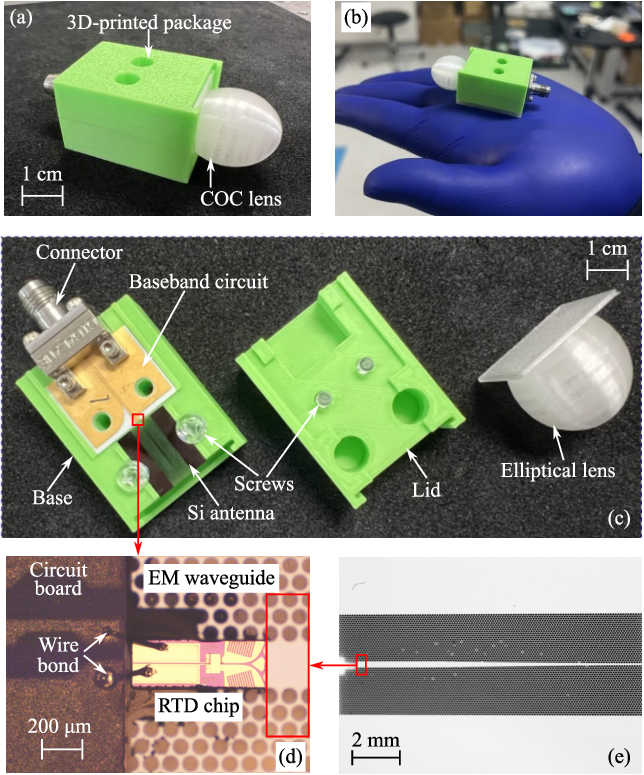}
	\caption{Realized RTD-based transceiver module. (a) Perspective view of the assembled module. (b) Module-on-palm demonstration highlighting its compactness. (c) Disassembled view of the module. (d) Microscope image of the waveguide-integrated RTD chip wire-bonded to baseband circuit. (e) Fabricated EM-clad tapered rod antenna.}
	\label{fig:realizedmodule}
\end{figure}
\begin{figure}[!tb]
	\centering
	\includegraphics[width=0.65\linewidth]{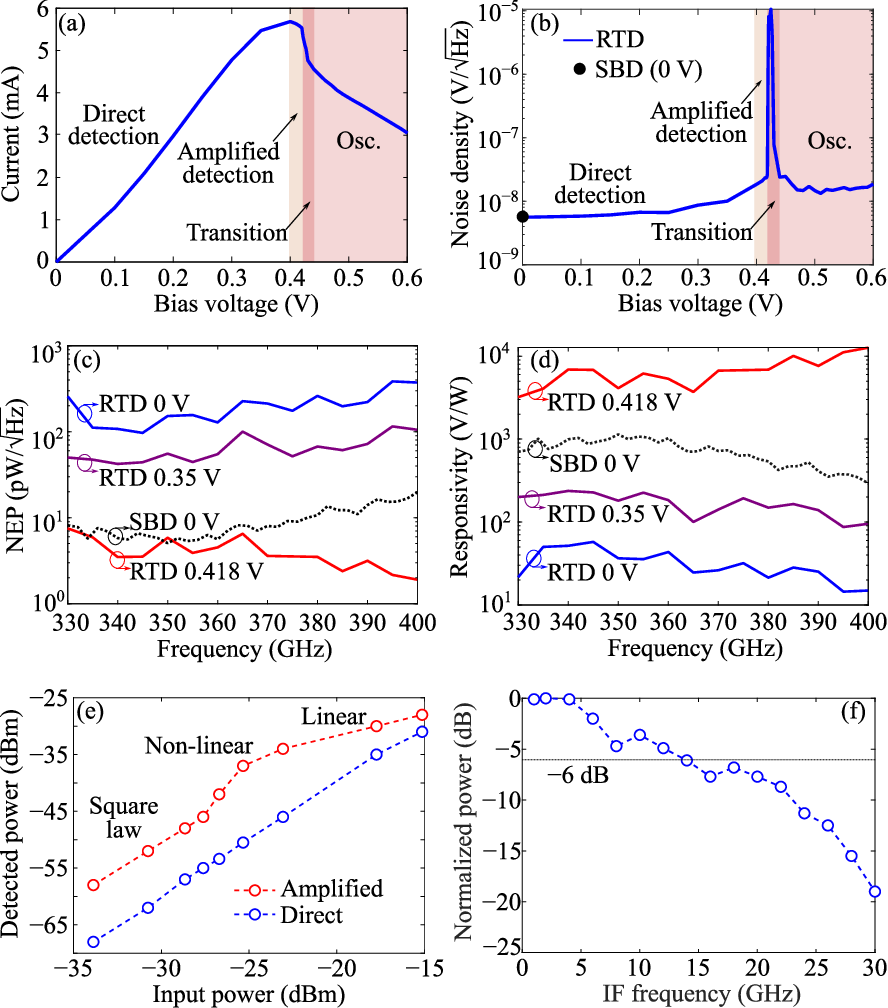}
	\caption{Characteristics of the RTD chip. (a) Measured current-voltage ($I$-$V$) characteristics with the effect of shunt resistor excluded. (b) Measured noise voltage density at vias bias voltages. (c) Measured NEP. (d) Measured responsivity. (e) Measured detected power of the RTD chip against input power for amplified detection and direct (square-law) detection at the carrier frequency of 300 GHz and an IF frequency of 1 GHz. (f) IF bandwidth of the RTD chip at the carrier frequency of 300 GHz with normalized output power. Osc.: Oscillation.}
	\label{fig:rtdchrac}
\end{figure}
The realized RTD-based transceiver module is shown in Figs.~\ref{fig:realizedmodule}(a)-(c). The module packaged in a 3D-printed PLA enclosure and equipped with a COC asymmetric elliptical lens. PLA is chosen for its cost-effectiveness, mechanical robustness, and tolerances to small fabrication errors~\cite{ichikawa2025}. The infill factors for the lens and the package are set to $100\%$ and $15\%$, respectively. The lower infill of the package is chosen to reduce overall weight while maintaining sufficiently mechanical strength. The package has a small air cavity inside to minimize unwanted wave interference from the package walls. The EM-clad tapered rod antenna together with the RTD chip and baseband circuit is mounted on the base. To mitigate dielectric loss introduced by the PLA package, a COC substrate is used to support the dielectric rod antenna and RTD chip. The RTD is fixed with glue and wire bonded to the baseband circuit as shown in Fig.~\ref{fig:realizedmodule}(d), which connects to a 2.92-mm coaxial connector for signal I/O. The dielectric antenna shown in Fig.~\ref{fig:realizedmodule}(e), is secured using two plastic screws. The top and bottom parts of the package are also fastened with plastic screws, while the COC lens, co-printed with a supporting layer, is directly inserted into the assembled structure. It is noticed that a air gap less than 1~mm remains between the lens and the rod antenna, but it has a negligible impact on antenna gain, as verified through simulations. 

The measured $I$–$V$ characteristic of the RTD chip is shown in Fig.~\ref{fig:rtdchrac}(a), revealing distinct regions corresponding to direct detection, amplified detection, and negative differential conductance (NDC) for oscillation. The noise voltage density as a function of bias is presented in Fig.~\ref{fig:rtdchrac}(b). In the direct detection region, the noise remains relatively constant, while in the amplified detection region it increases to $2.8\times10^{-8}$~$\rm{V}/\sqrt{\rm{Hz}}$ at 0.418~V from $5.6\times10^{-9}$~$\rm{V}/\sqrt{\rm{Hz}}$ at zero bias condition. A noise peak is observed near the transition region, while in the oscillation regime the noise stabilizes around $1.5\times10^{-8}$~$\rm{V}/\sqrt{\rm{Hz}}$. The corresponding noise equivalent power (NEP), shown in Fig.~\ref{fig:rtdchrac}(c), decreases with increasing bias, reaching an average value of around 4~$\rm{pW}/\sqrt{\rm{Hz}}$ across 330–400~GHz and a minimum of 1.8~$\rm{pW}/\sqrt{\rm{Hz}}$ at 400~GHz. This performance is comparable to state-of-the-art Fermi-level managed barrier diodes (FMBDs) \cite{ito2017inp} and shows increasing advantage over Schottky barrier diodes (SBDs) \cite{VDI_ZBD_2026}. The extracted responsivity, shown in Fig.~\ref{fig:rtdchrac}(d), reaches an average value of around 6.8~kV/W in the 300-GHz band under amplified detection, which is approximately an order of magnitude higher than SBDs. This enhanced sensitivity arises from the intrinsic nonlinear amplification mechanism of the RTD without triggering oscillation. The input–output relationship in Fig.~\ref{fig:rtdchrac}(e) further demonstrates the gain enhancement in the amplified detection regime. The measured IF response exhibits an approximate 15~GHz bandwidth, as shown in Fig.~\ref{fig:rtdchrac}(f), consistent with the 6-dB roll-off expected from square-law detection behavior. It should be noted that these measurements are performed using a ground-signal (GS) probe~\cite{headland2022rtd,yu2021hybrid,kawamoto2023rtd}, and the absence of wire-bonding effects may result in slightly overestimated IF bandwidth compared to fully packaged module.

\subsection{Wireless communications}
\subsubsection{Communications with RTD module as receiver}
\begin{figure}[!tb]
	\centering
	\includegraphics[width=0.75\linewidth]{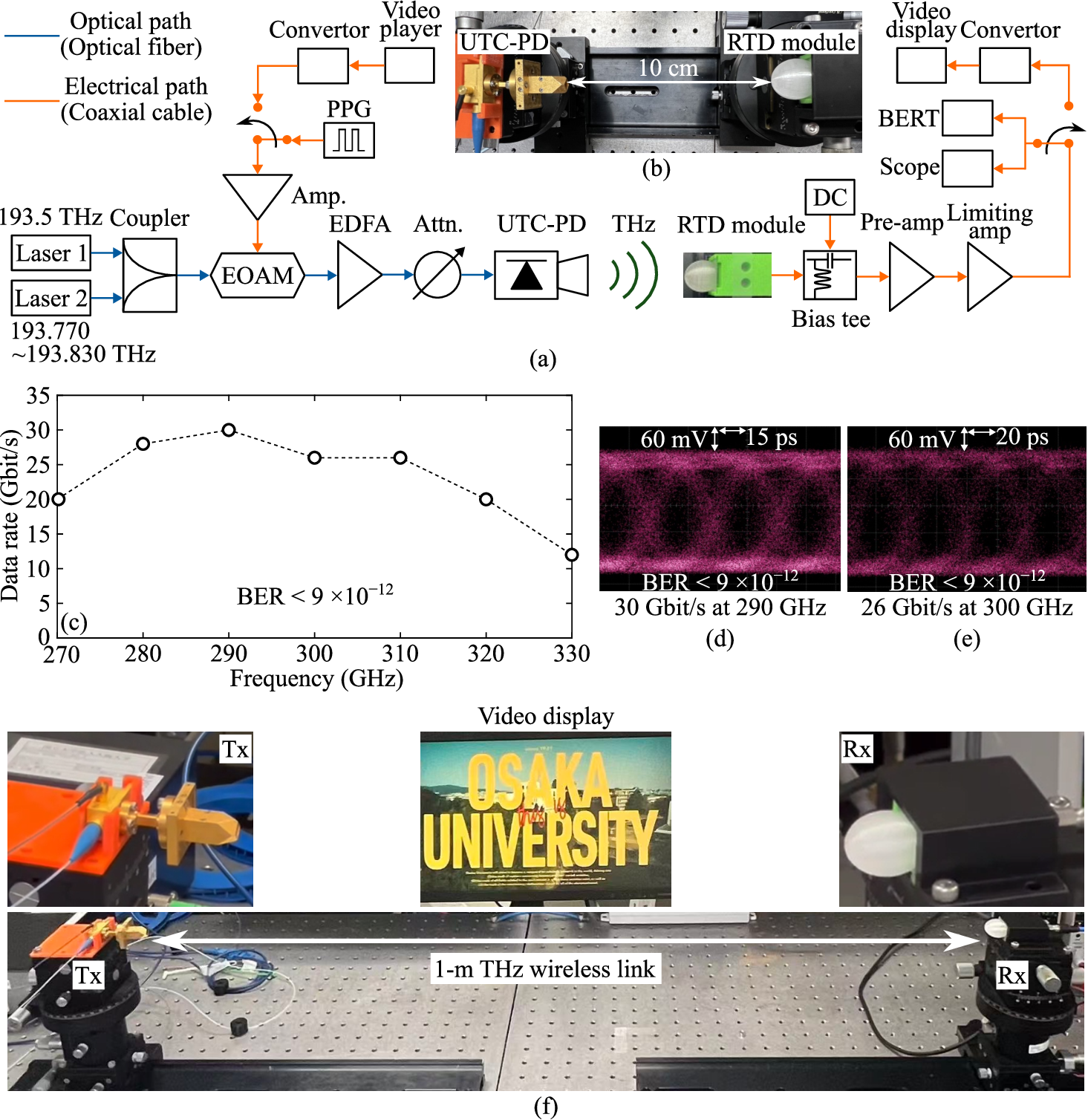}
	\caption{Wireless communications with the RTD module as a receiver. (a) Block diagram of the communications setup. (b) Photo of the terahertz wireless link. (c) Error-free transmission data rate against carrier frequency for OOK modulation. Eye-diagrams for error-free transmission at (d) 290~GHz and (e) 300~GHz. (f) Close-up view of terahertz link setup for 1-m HD video transmission. EOAM: Electro-optical amplitude modulator, Amp.: Amplifier, PPG: Pulse pattern generator, EDFA: Erbium-doped fiber amplifier, Attn.: Optical variable attenuator, BERT: Bit error rate tester.}
	\label{fig:commsetup}
\end{figure}

The wireless communications experiment adopting the RTD module as receiver is conducted using the setup shown in~Fig.~\ref{fig:commsetup}(a). On the transmitter side, a photonics-assisted setup is adopted, where two free-running lasers are employed to generate an optical beat signal in terahertz range. The combined optical signals are data modulated using an electroptic amplitude modulator (EOAM) driven by a pulse pattern generator, and then amplified by an erbium-doped fiber amplifier (EDFA). The amplified optical carriers are fed into a uni-traveling-carrier photodiode (UTC-PD) with a hollow-waveguide interface. The terahertz signal is generated through photomixing, and radiated from the horn antenna. On the receiver side, the RTD module retrieves the modulating data from the incoming terahertz signals, with the bias set in the amplified detection region. A pre-amplifier and limiting amplifier reshape the recovered signals for analysis using a bit error rate (BER) tester and oscilloscope.

A BER test is performed over a 10-cm wireless link as shown in Fig.~\ref{fig:commsetup}(b), with a UTC-PD output power of around 20~$\upmu$W and the carrier frequency varied from 270 to 330~GHz. The RTD receiver is operated at the amplified square-law detection region as shown in Fig.~\ref{fig:rtdchrac}(b). The maximum error-free (BER $<10^{-11}$) data rate of 30~Gbit/s is achieved at 290~GHz, slightly below the system limit of 32~Gbit/s for the on-off keying (OOK) modulation. The corresponding eye-diagrams at 290 and 300~GHz shown in Figs.~\ref{fig:commsetup}(d)-(e), exhibit clear openings, confirming stable signal recovery. The degraded performance at the lower and higher frequencies is primarily attributed to the roll-off effect of the UTC-PD and the limited IF bandwidth of the RTD. The wire-bonding between the RTD chip and the baseband circuit also impacts the IF response. Specifically, variations in bonding wire length and position can introduce parasitic inductance at low frequencies, thereby degrading the IF bandwidth of the module. Those effects are verified by simulations, which can be found in the Appendix. Although this effect can be qualitatively reproduced in simulations, precise control remains challenging with semi-automatic bonding and can be improved with finer bonding techniques. As shown in Fig.~\ref{fig:commsetup}(f), a real-time high-definition (HD) video transmission is successfully demonstrated over 1~m wireless link at the 300~GHz band, using a transmitting power approximately 20~$\upmu$W. The demonstration can be found from \href{https://drive.google.com/file/d/1ZS_GstR1Qm7dWkJgLiq_hMwkmuq-kl90/view?usp=sharing}{Visualization}. This experiment highlights the capability of the compact RTD module, which enables long-range terahertz communications without requiring external bulky optical lenses, marking an important step toward handheld, high-speed wireless systems. 

\begin{figure}[!htb]
	\centering
	\includegraphics[width=0.75\linewidth]{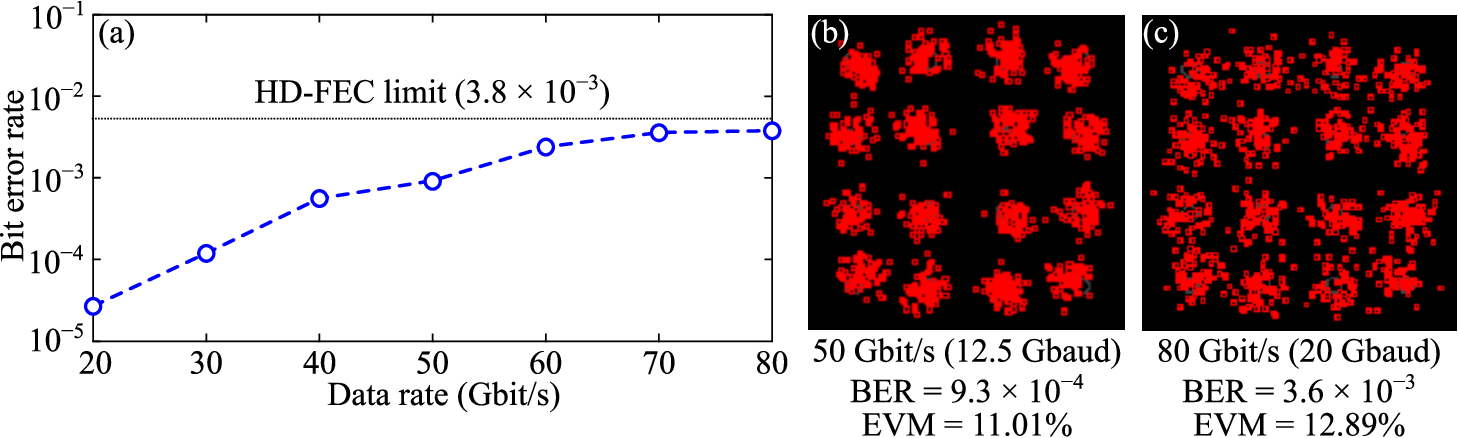}
	\caption{Communications performance with 16-QAM modulation. (a) BER against data rate. Constellation diagrams with data rate of (b) 50~Gbit/s and (c) 80~Gbit/s with BER below HD-FEC limit. The communications setup is similar with that shown in Fig.~\ref{fig:commsetup}, where the PPG is replaced with an AWG, while the BERT and oscilloscope are replaced with an RTO.}
	\label{fig:QAM}
\end{figure}

To further increase the data rate, 16-QAM IF transmission~\cite{ichikawa2025} is demonstrated over a 10-cm wireless link at 300~GHz, which lies at the center of the operation band and exhibits a broader 3-dB bandwidth, as shown in Fig.~\ref{fig:commsetup}(c). An arbitrary waveform generator (AWG) and real-time oscilloscope (RTO) are employed at the transmitter and receiver, respectively. With a transmitter power of 20~$\upmu$W and IF frequency of 12.5~GHz, the BER increases with data rate as shown in Fig.~\ref{fig:QAM}(a). The constellation diagram with a baud rate of 12.5~Gbaud is shown in Fig.~\ref{fig:QAM}(b), where the saturation is caused by RTD detector as shown in Fig.~\ref{fig:rtdchrac}(c). A maximum data rate of 80~Gbit/s (20~Gbaud) is achieved with BER below the threshold of hard-decision forward error correction (HD-FEC). The limitation is mainly set by the RTD's IF bandwidth. As illustrated in Figs.~\ref{fig:QAM}(b)-(c), the constellation diagram at a baud rate within the IF frequency exhibits significantly lower noise than when the baud rate exceeds the IF bandwidth. This degradation arises from signal-signal beat interference (SSBI), which introduces asymmetric IF spectra and nonlinear mixing between residual sidebands, leading to strong in-band signal interference~\cite{ichikawa2025}. Although the SSBI effects can be partially compensated by adaptive equalization in the RTO, further improvements are anticipated through optimization of the RTD circuit and wire-bonding interface. Compared with the state-of-the-art summarized in Table~\ref{tab:comm}, the proposed RTD module achieves both the longest transmission distance and highest data rate at the 300-GHz band, attributed to its broadband, high-efficiency antenna-lens integration strategy, while eliminating the package-induced metallic loss. These results clearly demonstrate the practicality of our technology, enabling a compact, low-cost, high-performance terahertz portable RF frontend.
\begin{table*}[!tb]
	\centering
	\caption{Comparison of terahertz communications adopting RTD as Tx and/or Rx (Tx:Transmitter, Rx: Receiver)}
	
	\begin{adjustbox}{width=1\textwidth}
		
		\begin{threeparttable}
			\begin{tabular}{lcccccccccc} \hline
				
				&\textbf{Frequency}&\textbf{Tx}&\textbf{Rx}&\textbf{Modulation}&\textbf{Detection}&\textbf{Tx $P_{\rm{out}}$}&\textbf{Distance}&\textbf{Total gain}\tnote{a}&\textbf{Data rate}&\textbf{BER}\\
				&(GHz)&&&\textbf{format}&\textbf{scheme}&(mW)&(mm)&(dBi)&(Gbit/s)\\ \hline
				
				\textbf{On-chip communications}&&&&&&&\\ 
				Unclad waveguide diplexer~\cite{yu2021hybrid} &297(Cross), 345(Bar)&UTC-PD&RTD&OOK&Direct&0.08&5&--&24\tnote{b}&$2.1\times10^{-12}$\\
				\\
				EM waveguide~\cite{ichikawa2025}&315&UTC-PD&RTD&32-QAM&Direct&0.02&50&--&100&$3.58\times10^{-3}$\\
				\\
				PC waveguide~\cite{yu2019}&350&UTC-PD&RTD&OOK&Direct&0.08&5&--&32&$2.1\times10^{-12}$\\
				\hline
				
				\textbf{Wireless communications}&&&&&&\\ 
				Bow-tie antenna~\cite{webber2023} &324&UTC-PD&RTD&16-QAM&Direct&0.04&10&52&68&$2\times10^{-3}$\\
				\\
				Bow-tie antenna~\cite{webber20218k} &324 (Ch1), 335 (Ch2)&UTC-PD&RTD&OOK&Coherent&0.03&20--30&52&48&$<2\times10^{-12}$\\
				\\
				Bow-tie antenna~\cite{oshiro2022} &343&UTC-PD&RTD&PAM4&Direct&0.02&5&52&48&$1.98\times10^{-3}$\\
				\\
				Bow-tie antenna~\cite{nishida2019} &348&RTD&RTD&OOK&Coherent&0.028&70&54&30&$1\times10^{-11}$\\
				\\
				Modified Vivaldi antenna~\cite{ngo2023rtd} &350&UTC-PD&RTD&OOK&Direct&0.02&30&50&25&$2.1\times10^{-12}$\\
				\\
				$2\times2$ slot antenna array~\cite{oshima2017} &790&RTD&SBD&OOK&Direct&0.03&200\tnote{c}&84&28\tnote{d}&$1.5\times10^{-3}$\\
				\\
				Cavity-ring-slot antenna array~\cite{ngo2025} &860&RTD&RTD&OOK&Coherent&0.02&5&20&1.2&$1\times10^{-9}$\\
				\\
				\textbf{This work}&290&UTC-PD&RTD&OOK&Direct&0.02&1000&55&1.5&$<9\times10^{-12}$\\
				\textbf{(Lens-coupled rod antenna)}&290&UTC-PD&RTD&OOK&Direct&0.02&100&55&30&$9\times10^{-12}$\\
				&300&UTC-PD&RTD&16-QAM&Direct&0.02&100&55&80&$3.6\times10^{-3}$\\
				&332&RTD&RTD&OOK&Direct&0.005&30&60&20&$3\times10^{-5}$\\
				
				\hline
				
			\end{tabular}
			\begin{tablenotes}
				\item[a]Total antenna/lens gains with Tx and Rx combined.
				\item[b]Maximum aggregated error-free data rate with dual-channel simultaneous transmission.  
				\item[c]External lenses are adopted.
				\item[d]Single channel data rate.
			\end{tablenotes}
		\end{threeparttable}
	\end{adjustbox}	
	\label {tab:comm}
\end{table*}
%\threesubsection{First lowest-level subsection}
\subsubsection{Communications with RTD module as transmitter}
\begin{figure}[!tb]
	\centering
	\includegraphics[width=0.75\linewidth]{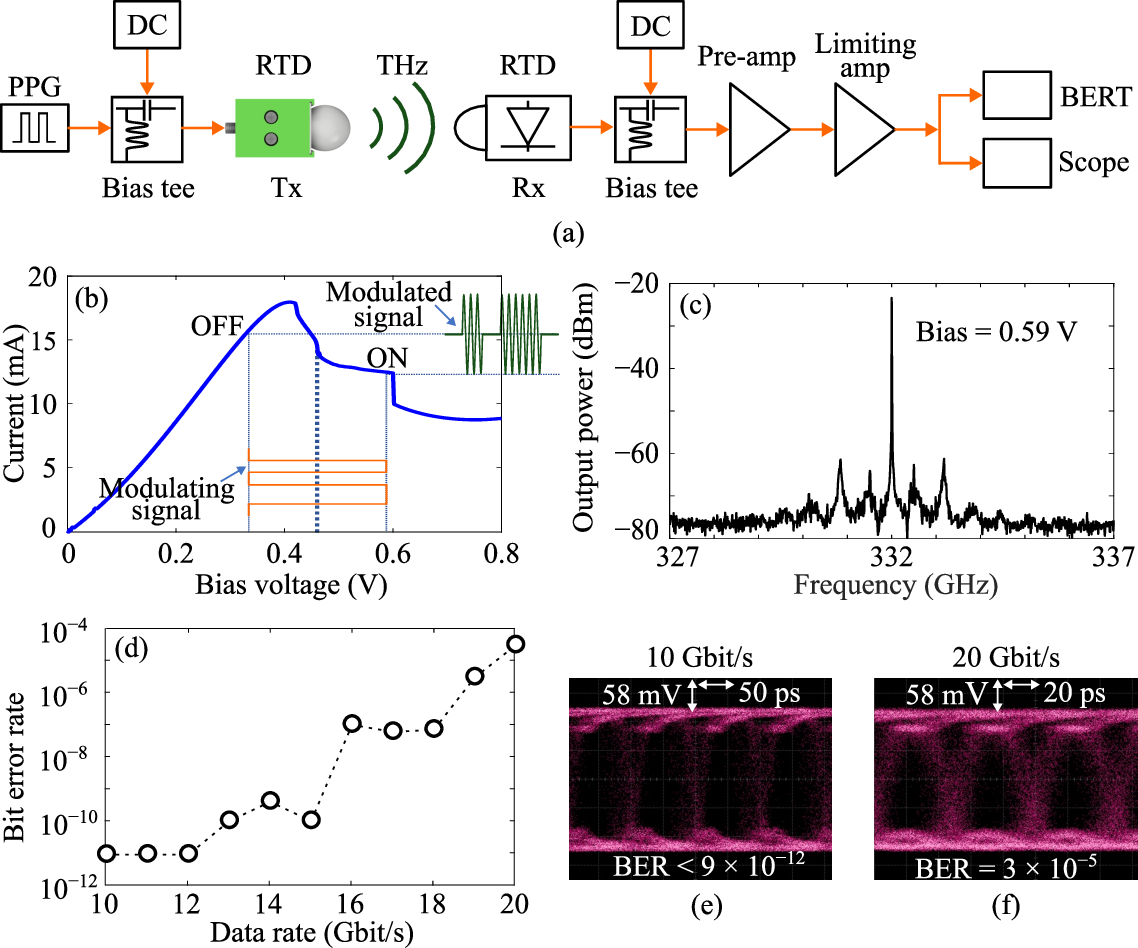}
	\caption{Wireless communications with the RTD module as a transmitter. (a) Block diagram of communications setup. (b) OOK modulation scheme based on the measured current-voltage ($I$-$V$) curve. (c) Measured oscillation frequency spectrum without modulation at bias voltage of 0.590~V. (d) BER at the carrier frequency of 332~GHz. Eye diagrams for OOK transmission at data rate of (e) 10~Gbit/s and (f) 20~Gbit/s. PPG: Pulse pattern generator, DC: Direct current source, Pre-amp, Pre-amplifier, Limiting amp: Limiting amplifier, BERT: Bit error rate tester, Scope: Oscilloscope.}
	\label{fig:Tx}
\end{figure}
Wireless communications experiments with the RTD module operating as a transmitter are conducted using the setup shown in Fig.~\ref{fig:Tx}(a). At the transmitter side, a pulse pattern generator provides the modulating signal, which is combined with the DC bias through a bias tee and applied to the RTD module. At the receiver side, a metallic RTD module integrated with a COC lens antenna is employed as a detector. The received signals are subsequently amplified and reshaped using a pre-amplifier with a 3-dB bandwidth of 38~GHz and a limiting amplifier operated for up to 45~Gbit/s, followed by a bit error tester and an oscilloscope for performance evaluation. Compared to photonics-based systems, this fully electronic configuration offers a significantly simpler architecture. OOK modulation is realized by varying the bias voltage of the RTD. As illustrated in Fig.~\ref{fig:Tx}(b), a DC bias of 0.465~V is initially applied to sustain terahertz oscillation. A peak-to-peak modulating voltage of 0.25~V is then introduced, switching the RTD between the non-oscillating state at 0.340~V (off state) and the oscillating state at 0.590~V (on state). This enables OOK modulation of the emitted terahertz signal at a carrier frequency of 332~GHz, as shown in Fig.~\ref{fig:Tx}(c). The output spectrum shown in Fig.~\ref{fig:Tx}(c) is measured using a spectrum analyzer extender (SAX) (Virginia Diodes Inc.) connected to a signal analyzer. The wireless distance between the transmitter and the SAX (equipped with a horn antenna) is approximately 3~cm, consistent with the link distance used in the communications experiment. The measured carrier power is approximately $-23$ dBm, which corresponds to the received power level in the communications setup.
The BER performance as a function of data rate is presented in Fig.~\ref{fig:Tx}(d). Error-free transmission is achieved up to 12~Gbit/s, while the BER increases to $3\times10^{-4}$ at 20~Gbit/s. The corresponding eye diagrams at 10~Gbit/s and~20 Gbit/s are shown in Figs.~\ref{fig:Tx}(e)-(f), respectively. The data rate is currently limited by the relatively low output power of the RTD, which can be improved through optimization of the RTD mesa dimensions and circuit design~\cite{diebold2016}. These results experimentally confirm that the proposed module can operate as both a transmitter and a receiver, thereby validating its functionality as a terahertz transceiver above 300~GHz.

\section{Conclusion}
We have proposed an RTD-enabled terahertz wireless transceiver module based on electronic-photonic antenna integration. A modified metallic Vivaldi with enhanced bandwidth and directivity is integrated with the RTD, serving as a broadband mode converter to the dielectric waveguide with high coupling efficiency. The interconnection between a dielectric tapered rod antenna and 3D-printed COC elliptical lens achieves an average gain above 30~dBi over 220--330~GHz for both orthogonal polarizations, while an EM-clad tapered antenna combined with an asymmetric elliptical lens can further enhance the mechanical robustness without compromising performance. A 3D-printed package replaces conventional metallic package, significantly reducing cost and minimizing package-induced losses. Operated as a receiver, wireless communications demonstrate a maximum error-free data rate of 30~Gbit/s with OOK modulation and 80~Gbit/s with 16-QAM modulation with BER below HD-FEC limit at the 300 GHz band. In addition, real-time HD video transmission over 1~m validates the system's practical performance. Moreover, error-free transmission with a data rate up to 12~Gbit/s (OOK) is achieved with the module adopted as a transmitter. These results demonstrate the feasibility of the proposed transceiver module for real-world terahertz applications. The module can be further expanded into a multifunctional frontend through the monolithic integration of various passive components on the EM-clad waveguide platform. It can be foreseen that such a hybrid integration technology combining silicon photonics, electronic/photonic devices, and 3D printing, is expected to enable compact, low-cost, and high-performance portable terahertz systems for future 6G and beyond.

% Experimental section

\section{Experimental Section}
\subsection{Lens antenna radiation patterns measurement}
\begin{figure}[!h]
	\centering
	\includegraphics[width=0.7\linewidth]{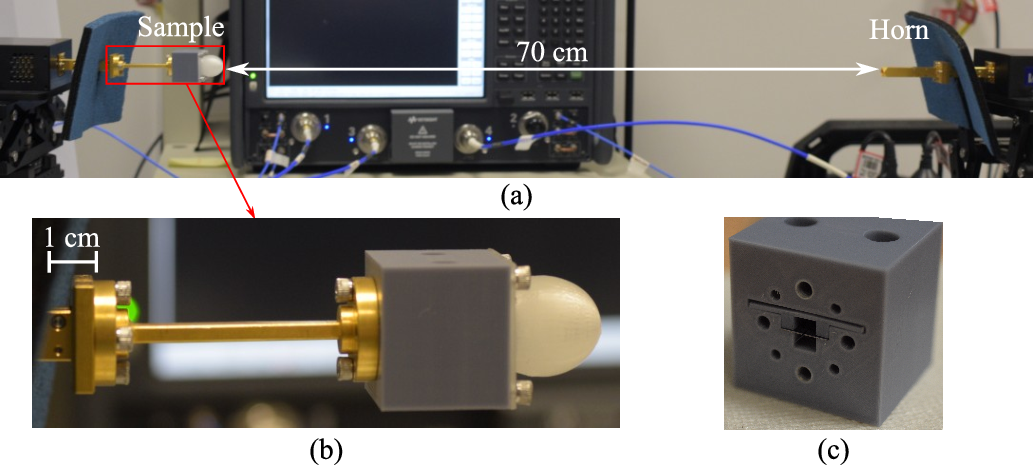}
	\caption{Radiation patterns measurement for the air-clad rod antenna coupled to a 3D-printed elliptical lens. (a) Overview of the measurement setup with a distance of 70~cm. (b) Magnified view of the assembled antenna connected to a straight hollow waveguide for the $E_{11}^x$ mode measurement. (c) Backside view of the package with a WR-3.4 flange. (d) Magnified view of the tapered structure of the dielectric waveguide.  The rotation angle for the setup ranges from $\pm75^{\circ}$. A twisted waveguide is used to replace the straight hollow waveguide to measure cross-polarization components and the $E_{11}^y$ mode with orthogonal polarization. Tx: Transmitter, Rx: Receiver.}
	\label{fig:rpsetup}
\end{figure}

The silicon antenna is fabricated based on a deep reactive-ion etching (DRIE) process on a 200-$\upmu$m high-resistivity ($>$ 10 k$\Omega\cdot$cm) intrinsic float-zone silicon wafer. The detailed fabrication procedure is provided in~\cite{gao2025ultra}. The RTD chip is fabricated with a process involving lithography and wet etching at ROHM Semiconductor Co. Ltd, with additional details available in~\cite{yu2019,yu2021hybrid}. The 3D printing is performed using a Bambulab A1 mini printer with a nozzle size of 0.2 mm and layer height of 0.05 mm. CREMELT COC and eSUN PLA+ are used for printing the elliptical lens and packages, respectively. The reflection coefficient and antenna radiation pattern measurements are performed using a Keysight vector network analyzer (VNA) with VDI WR-3.4 extension mudules spanning from 220-330~GHz as shown in Fig.~\ref{fig:rpsetup}. The air-clad tapered rod antenna, fed by the EM waveguide, is housed in a 3D-printed package designed with a WR-3 waveguide flange to ensure a robust interface with the feeding hollow waveguide. The coupling taper of the EM waveguide is inserted into the hollow waveguide, while a metallic horn antenna is positioned at the receiver side at a distance of approximately 70~cm, satisfying far-field 
measurement conditions.
\subsection{RTD sensitivity and noise measurement}
\begin{figure}[!htb]
	\centering
	\includegraphics[width=0.7\linewidth]{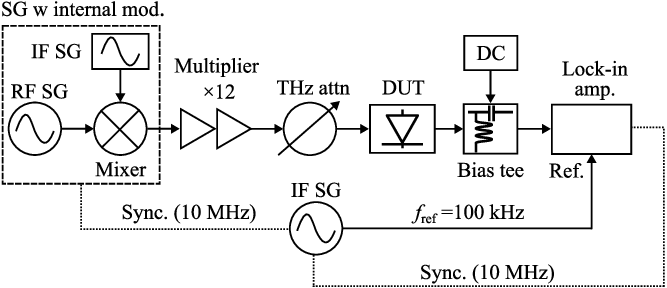}
	\caption{Block diagram of the measurement setup for RTD noise density, noise equivalent power, and responsivity. The IF frequency is 100 kHz equal to the reference frequency of the lock-in amplifier.}
	\label{fig:NEPsetup}
\end{figure}
\begin{figure}[!tb]
	\centering
	\includegraphics[width=0.48\linewidth]{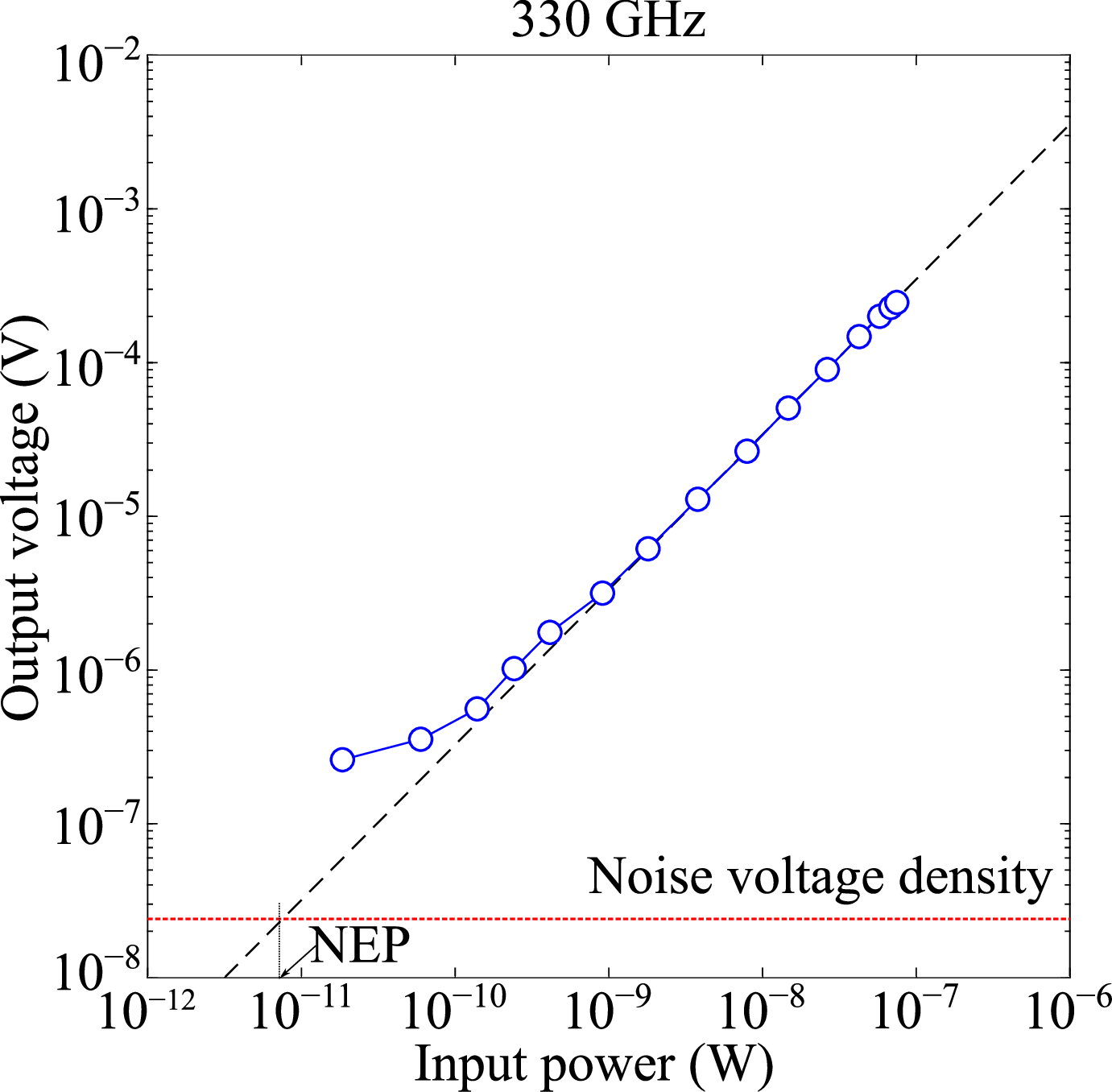}
	\caption{Relationships between input power and output voltage for RTD as a receiver at 330 GHz. The measured noise voltage density (V/$\sqrt{\rm{Hz}}$) is shown by the dashed line with the bias voltage of 0.418 V (amplified detection). The intersection between the fitted curve and the noise voltage density corresponds to NEP.}
	\label{fig:NEP330}
\end{figure}

To evaluate the sensitivity and noise performance of the RTD module as a receiver, a terahertz electronic measurement setup is employed, as shown in Fig.~\ref{fig:NEPsetup}. A signal generator produces a millimeter-wave signal (27.5–33.3 GHz) modulated at 100 kHz, which is upconverted to 330–400 GHz using a $\times$12 multiplier chain. Variable attenuators are used to reduce the incident terahertz power down to picowatt level. The RTD module is biased via a bias tee, and the demodulated signal is measured using a lock-in amplifier (NF Corporation LI5660)~\cite{NF_Lockin_2026} referenced to the 100 kHz modulating frequency generated by a separated signal generator. All instruments are synchronized to a common 10~MHz reference to enhance the of measurement sensitivity. The noise voltage density is measured using the built-in noise function of the lock-in amplifier with no input signal. The NEP is then estimated by measuring the output voltage as a function of input power and extrapolating the linear response region, as shown in Fig.~\ref{fig:NEP330}. The NEP is defined as the input power per unit frequency bandwidth when the SNR is 1. When output voltage and noise voltage density are plotted on the same graph, NEP is estimated from the input power at which the two lines intersect. Experimentally, it is observed that when the input power is reduced to around $6\times10^{-11}$~W, the output voltage tends to saturate and remains nearly constant with further decreases in input power. Although this effect can be partially mitigated by extending the measurement time, the true NEP cannot be directly measured due to the sensitivity limitation of the lock-in amplifier. Therefore, a linear curve fitting approach is applied to estimate the NEP, as illustrated in Fig.~\ref{fig:NEP330}. The responsivity is subsequently calculated as the ratio between output voltage and input power, equivalently expressed as the ratio of noise voltage density to NEP. In this characterization, the RTD chip is packaged in a metallic module with a WR-2.2 hollow waveguide interface~\cite{ngo2023rtd} to facilitate coupling with the transmitter and to match the operating frequency range of the available measurement equipment. Accordingly, the performance metrics are characterized over 330–400 GHz. Given that the RTD module is designed for ultra-broadband operation, as detailed in Section 2, this frequency range provides a representative evaluation of its overall performance.
\subsection{RTD detected power measurement}
\begin{figure}[!tb]
	\centering
	\includegraphics[width=0.7\linewidth]{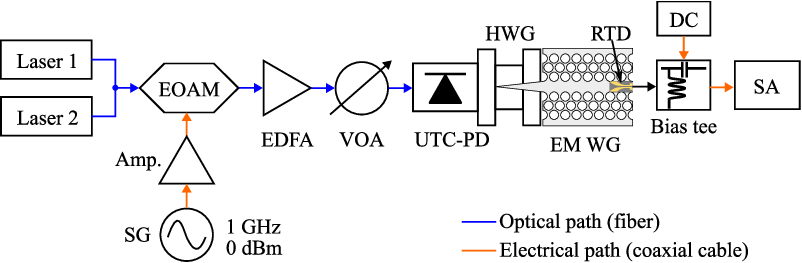}
	\caption{Block diagram of measurement setup for RTD detected power. SG: Signal generator, EOAM: Electro-optical amplitude modulator, EDFA: Erbium-doped fiber amplifier, VOA: Variable optical attenuator, HWG: Hollow waveguide, SA: Signal analyzer.}
	\label{fig:powermea}
\end{figure}
The detected power shown in Fig.~\ref{fig:rtdchrac}(e) corresponds to the electrical power of the demodulated IF signal at 1~GHz, measured using a signal analyzer with a 50~$\ohm$ input impedance. The measurement setup is illustrated in Fig.~\ref{fig:powermea}. At the transmitter side, two optical tones with a frequency spacing of 300~GHz are intensity-modulated by a 1 GHz sinusoidal signal (0~dBm) and injected into a UTC-PD for terahertz generation. A variable optical attenuator is used to control the emitted terahertz power. The terahertz signal is coupled from the UTC-PD module into the EM-clad waveguide via a tapered transition, and subsequently coupled to the RTD chip functioning as a receiver. The RTD is biased at 0.350~V and 0.418~V, corresponding to direct and amplified detection regimes, respectively. The resulting IF signal is measured by the signal analyzer. It should be noted that the coupling losses between the dielectric waveguide and the hollow waveguide, as well as between the waveguide and the RTD, are not de-embedded. Therefore, the actual terahertz power incident on the RTD is slightly lower than the reported value.

\section*{Appendix}
\begin{figure}[!b]
	\centering
	\includegraphics[width=0.7\linewidth]{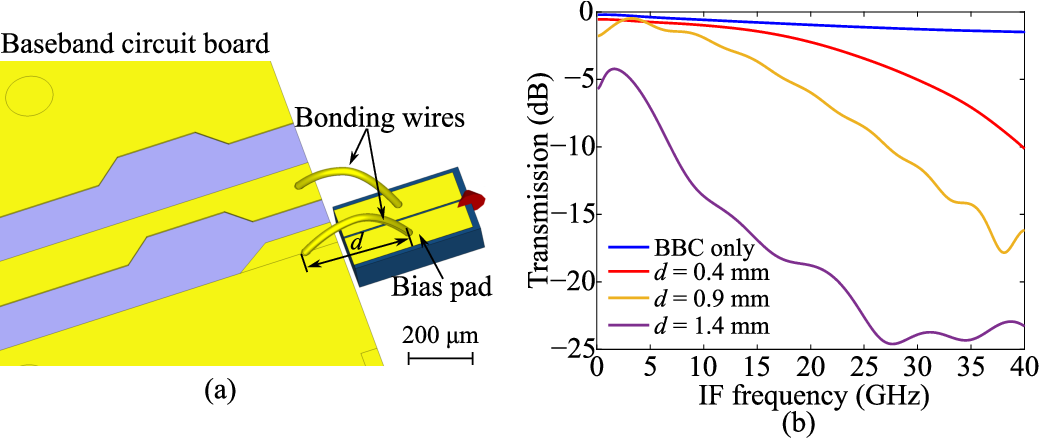}
	\caption{Impact of the wire bonding on the IF bandwidth of the RTD module. (a) Magnified view of the interconnection between the RTD chip and the baseband circuit through wire bonding. (b) Simulated IF bandwidth with various bonding wire lengths.}
	\label{fig:bbc}
\end{figure}
To investigate the impact of bonding wires on the IF bandwidth, we analyze the transmission between the baseband circuit and the RTD bias pad, as shown in Fig.~\ref{fig:bbc}(a). The wire length is parameterized by the projected distance $d$, where a larger $d$ corresponds to a longer wire and increased separation from the baseband circuit. To single out the effect of the interconnect, the RTD is modeled as a discrete port in CST simulations. As shown in Fig.~\ref{fig:bbc}(b), in the absence of bonding wires, a 3-dB IF bandwidth of the baseband circuit board exceeding 40~GHz can be achieved, indicating that the intrinsic bandwidth of the system is sufficiently large. However, when bonding wires are introduced to interconnect the baseband circuit board and the RTD bias pads, the IF bandwidth decreases significantly, accompanied by increased insertion loss. This degradation becomes more pronounced with increasing wire length. For example, at $d=1.4$~mm, the 6-dB IF bandwidth is reduced to approximately 10~GHz, with an insertion loss exceeding 4~dB. These results indicate that the IF bandwidth of the packaged module is highly sensitive to the parasitic inductance and impedance mismatch introduced by wire bonding. Therefore, precise control of wire length and placement, or alternatively the adoption of advanced interconnect techniques (e.g., flip-chip integration), is critical for achieving broadband operation.

%\threesubsection{First part of experimental section}\\
%\threesubsection{Second part of experimental section}\\

\medskip
\textbf{Supporting Information} \par %Please delete the Suppporting Information statement if it is not applicable. Please supply Supporting Information in another file. Supporting information should not be provided in .tex format
The demonstration of 1-m HD-video transmission is available at \href{https://drive.google.com/file/d/1ZS_GstR1Qm7dWkJgLiq_hMwkmuq-kl90/view?usp=sharing}{Visualization}.

% Acknowledgements
\medskip
\textbf{Acknowledgments} \par %delete if not applicable))
The authors would like to acknowledge Dr. Hiroshi Ito (the University of Tokyo) for his valuable suggestions regarding the NEP measurements. This work was supported in part by the Core Research for Evolutional Science and
Technology (CREST) program of the Japan Science and Technology Agency
(No. JPMJCR21C4) and KAKENHI (24H0003).

% References
\medskip
\bibliographystyle{IEEEtran}
\bibliography{Bibliography}

\end{document}